\documentclass[manuscript,nonacm]{acmart}
\usepackage{xcolor}
\usepackage[ruled,vlined,linesnumbered]{algorithm2e}

\AtBeginDocument{}
\setcopyright{none}
\acmDOI{}
\acmISBN{}

\usepackage[nolist]{acronym}
\begin{acronym}
    \acro{HCI}{Human-Computer Interaction}
    \acro{CSCW}{Computer-Supported Collaborative Work}
    \acro{CS}{Computer Science}
    \acro{LLM}{Large Language Model}
    \acro{AST}{Abstract Syntax Tree}
    \acro{AoI}{Area of Interest}
    \acro{LMM}{Linear Mixed Model}
    \acro{GLMM}{General Linear Mixed Model}
    \acro{EMM}{Estimated Marginal Mean}
    \acro{ML}{Machine Learning}
\end{acronym}
\usepackage{dirtytalk}
\usepackage{makecell}
\usepackage{indentfirst}
\usepackage{graphicx}
\usepackage{subcaption}
\usepackage{caption}
\usepackage{booktabs}
\usepackage{multirow}
\usepackage{soul}  
\usepackage{siunitx}
\usepackage{xkeyval}
\usepackage{wrapfig}
\author{Xiaotian Su}
\orcid{0009-0004-0548-1576}
\affiliation{%
  \institution{ETH Zurich}
  \city{Zurich}
  \country{Switzerland}}
\email{xiaotian.su@inf.ethz.ch}

\author{Jan Brasser}
\affiliation{%
  \institution{University of Zurich}
  \city{Zurich}
  \country{Switzerland}
}
\email{jan.brasser@uzh.ch}

\author{David Robert Reich}
\affiliation{%
  \institution{University of Zurich}
  \city{Zurich}
  \country{Switzerland}
}
\email{davidrobert.reich@uzh.ch}

\author{Lena A. Jäger}
\affiliation{%
  \institution{University of Zurich}
  \city{Zurich}
  \country{Switzerland}}
\email{lenaann.jaeger@uzh.ch}

\author{April Yi Wang}
\orcid{0000-0001-8724-4662}
\affiliation{%
  \institution{ETH Zurich}
  \city{Zurich}
  \country{Switzerland}}
\email{april.wang@inf.ethz.ch}

\renewcommand{\shortauthors}{Su et al.}

\begin{document}
\newcommand{\AW}[1]{\textcolor{blue}{\textbf{*April*}: #1}}
\newcommand{\XS}[1]{\textcolor{purple}{\textbf{*Xiaotian*}: #1}}
\newcommand{\tmp}[1]{\hl{#1}\xspace}

\newcommand{\tmpcite}{\tmp{[cite]}}
\newcommand{\psay}[1]{\textit{\say{#1}}}

\begin{abstract}
AI coding interfaces present generated code either all at once or token-by-token. These rendering strategies reflect model generation rather than how programmers actually read code: selectively, non-linearly, and guided by the structure. We argue that code rendering is a first-class interaction primitive that shapes how programmers read and understand code. To explore this design space, we introduce structured rendering, a technique that reveals code in semantically meaningful chunks derived from its syntactic hierarchy, exposing high-level structure before low-level details. To isolate rendering effects on visual attention, we conducted an eye-tracking study with 53 participants comparing static, character-based, and structured rendering. Our findings show that rendering alters visual attention and reading behavior: dynamic rendering induces fewer but longer fixations and more sustained focus, while structured rendering further guides attention toward semantically meaningful units and supports high-level understanding. 
We release the anonymized dataset with an interactive demo to support future research.
\end{abstract}

\newcommand{\sys}{CodeGaze}

\title{From Tokens to Structure: Investigating Visual Attention in Progressive Code Rendering}





\begin{teaserfigure}
    \centering
    \includegraphics[width=\linewidth]{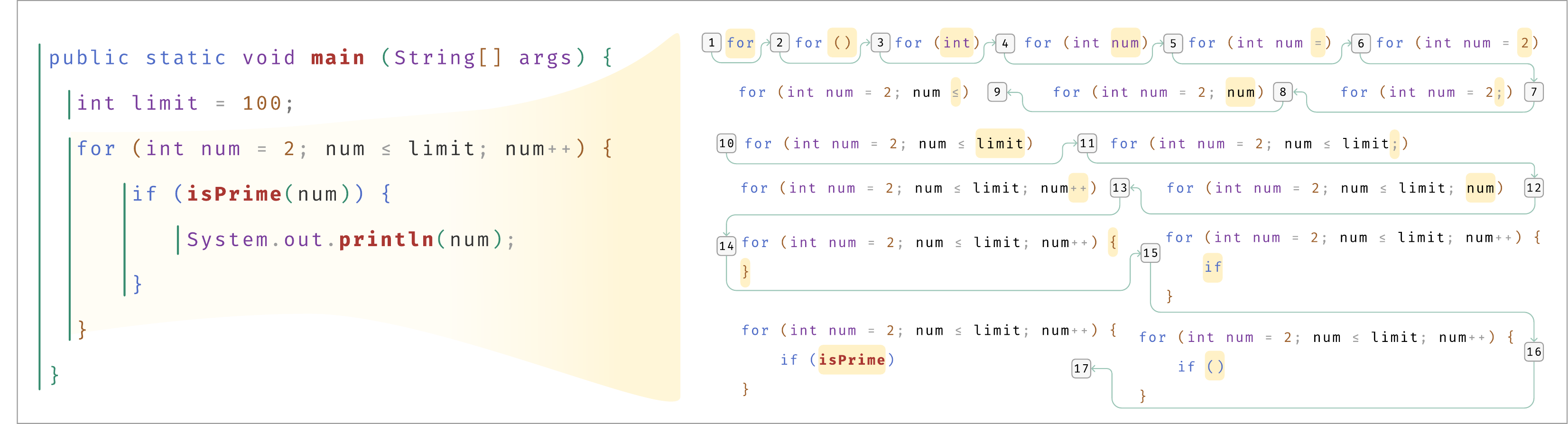}{
        \caption{Left: the main function of a Java program designed to find prime numbers up to a limit of 100. Vertical green rails mark \ac{AST} nesting depth (outer blocks to inner statements). Right: the dynamic process of structured rendering of the \texttt{for} loop, reveals its components in numbered steps that follow the syntactic hierarchy.}
        \Description{Left panel shows complete Java code for finding prime numbers with vertical green bars indicating Abstract Syntax Tree nesting levels. Right panel demonstrates six-step structured rendering process of a for loop: (1) skeleton "for ()", (2) initialization "int num = 2", (3) condition "num <= limit", (4) increment "num++", (5) body skeleton with braces, (6) complete inner if statement for prime checking.}
        \label{fig:render}
    }
\end{teaserfigure}
\maketitle

\section{Introduction}
Understanding source code is a core and cognitively demanding activity in software development, requiring programmers to construct mental models of program behavior in order to debug, modify, and extend systems~\cite{icpc15-relax, icse14-automated}.
With the rise of AI coding assistants, developers are increasingly required to read and understand generated code, making code comprehension more critical than ever.
However, current code presentation formats often follow model generation processes, revealing code in the speed and order it is produced rather than in ways optimized for human comprehension~\cite{uist25-streaming}.
While prior work has studied code comprehension and visual attention on static code, little is known about how dynamically rendered code affects how programmers allocate attention and construct understanding.

This gap matters because expert programmers do not read code in a uniform left-to-right fashion.
Prior work shows expert programmers read code in a non-linear, selective manner, guided by semantic cues and mental models rather than a strict left-to-right sequence~\cite{icpc15-relax, icse14-automated}. 
Rather than devoting equal attention to every token, experts strategically focus on semantic cues~\cite{fse17-neural}, and pay more attention to method signatures than the method body~\cite{icse14-automated}. 
This reading behavior enables programmers to perceive complex patterns holistically and rapidly extract higher-level abstractions from source code~\cite{shneiderman1979syntactic, fse17-neural}. 

Yet current AI interfaces for code generation typically present output either incrementally in generation order or all at once as complete suggestions.
In progressive rendering, code is revealed in a character-by-character sequence to mirror language model generation, as in systems such as ChatGPT and CodeAid~\cite{chi24-codeaid}. 
However, this approach does not align with the structural and semantic cues programmers rely on during comprehension.
In contrast, in static rendering, code is displayed as complete code blocks or as inline suggestions, as implemented in tools such as Claude Code\footnote{https://www.anthropic.com/claude-code} and Cursor AI\footnote{https://www.cursor.com/}.
These full-block suggestions can overload readers when outputs are lengthy and make selective inspection more difficult.

Therefore, rendering strategies that enforce generation order or uniform exposure may privilege the wrong units of attention.
We argue that rendering is not a passive wrapper around code output, but an active organizer of program comprehension.
This view is consistent with prior work on structured editing, which shows that editing interfaces aligned with program structure can reduce cognitive load and better support programmers’ mental models of code~\cite{chi23-structured, ieee23-structure}. 
Despite this, rendering remains largely undertheorized and under-evaluated. 
In this paper, we treat code rendering as a first-class interaction design variable and examine how different rendering strategies shape visual attention, navigation, and cognitive experience during code reading.
We frame the design space of code rendering along two dimensions: 
(1) \textbf{Static vs. Dynamic}: How does showing code all at once versus revealing it dynamically shape visual attention?
(2) \textbf{Structured vs Character-based}: If revealed dynamically, how does following the program’s \ac{AST} guide attention differently than a linear sequence?
Within this space, we introduce \emph{structured rendering}, a technique that renders code in semantically meaningful units.
The skeleton of a program (e.g., control-flow blocks, function headers) is displayed first, followed by incremental details. 
By leveraging the \ac{AST}, structured rendering seeks to align presentation with the hierarchical organization that programmers naturally use in comprehension. 

To examine how rendering styles affect visual attention, we conducted an eye-tracking study of dynamically revealed code ($N=53$). 
Using a high-resolution eye tracker, we capture fine-grained fixation and saccade patterns that allows us to compare how attention shifts across code tokens.
Compared with static code, dynamic rendering elicited fewer but longer fixations, suggesting differences in the allocation of attention.
Within the dynamic condition, participants in structured rendering, had longer and more variance in fixations, which may indicate deeper processing of the code and selective attention.
Participants also reported that structured rendering helped them see the scope first and better understand the overall structure.
These findings carry implications for both the design of human-centered AI coding interfaces and programming education.
In education, structured rendering can inform live coding demonstrations and video tutorials that aim to scaffold learners’ comprehension. 
For professional tools, aligning code presentation with program structure can help developers more quickly understand AI-generated code and reduce the risk of overlooking errors~\cite{liu2022unified, ase21-ast-transformer}.
To illustrate how these insights can be applied in practice, we further develop a VS Code extension\footnote{The current executable version of the Visual Studio Code extension is publicly available via the Open Science Framework (OSF). A finalized version will be released on the Visual Studio Code Marketplace upon acceptance of this paper.} that enables interactive control over when and how code is revealed. 
This prototype demonstrates how rendering can be treated as a user-controllable interaction, allowing developers to adapt code presentation to their needs.

In summary, this work makes three contributions: 
(1) We introduce structured rendering, an \acp{AST}-based technique for revealing code in semantically meaningful units, establishing rendering as a cognitive-aware design dimension for AI coding interfaces.
(2) We conduct a within-subject eye-tracking study (N=53) comparing static, character-based, and structured rendering. We release an anonymized dataset to support reproducible research on code presentation and visual attention.
(3) We provide empirical evidence and design insights showing how different code presentation formats influence programmers’ visual attention and reading strategies, informing the design of future AI-assisted programming interfaces.

\section{Related Work}
\label{sec:rw}
\subsection{Cognitive Process in Program Comprehension}
Program comprehension is a fundamental cognitive challenge in software development, defined as the process of constructing a mental model of a program being studied \cite{journal91-models}.
Unlike natural text comprehension, understanding source code requires reasoning about its unique structural and semantic properties, while drawing on strategies, prior experience, domain knowledge, and program knowledge \cite{journal12-expertise}.
A key mechanism in this process is chunking, through which programmers organize groups of statements into meaningful higher-level units rather than processing code symbol by symbol \cite{miller1956magical, shneiderman1979syntactic}. 
These units are then integrated hierarchically to form a broader understanding of the program, consistent with bottom-up models of program comprehension \cite{ieee16-prog-comp}. 

This perspective also helps explain why structured programs are often easier to understand and modify: their hierarchical organization better aligns with programmers’ cognitive organization \cite{MCKEITHEN1981307, shneiderman1979syntactic}. 
Recent tools on structured editing likewise aim to support chunk-based understanding \cite{chi23-structured, ieee23-structure}.
These cognitive accounts are also relevant for AI system design. 
Studies show that aligning \acp{LLM} with human attention can improve code understanding and generation by bringing model attention closer to human cognitive focus \cite{sfe24-attention, sfe24-eyetrans, tian2025spa}. 
Therefore, studying and understanding how programmers understand code is not only important for improving developer performance but also informing \ac{ML} researchers to design attention mechanisms that mirror human cognitive processes, thereby improving code generation models.

\subsection{Eye-Tracking for Program Comprehension}
Eye-tracking is widely used in software engineering research to study how programmers allocate visual attention while reading code~\cite{sharafi2015eye, etra12-scan, icpc12-women, icpc22-estimate}. 
By capturing visual behavior, it provides insight into the underlying cognitive processes~\cite{icpc22-estimate}.
Common metrics include fixation duration, fixation count, saccade amplitude, etc. 
Prior studies using these metrics have revealed systematic differences between novice and expert programmers, reflected in distinct eye-movement patterns.
However, the interpretation of these measures is often not one-to-one: the same pattern can reflect different underlying processes depending on the task and context.
\textbf{(1) Fixations and saccades.}
Evidence is mixed: some studies report that experts exhibit larger saccade amplitudes and shorter fixation durations than novices, suggesting more efficient visual processing \cite{andrzejewska2020development, icpc15-relax}, whereas others find that experts produce more fixations in tasks such as code reuse \cite{jessup2021using}.
\textbf{(2) Reading linearity.} 
A consistent finding is that experts read code less linearly than novices. 
Novices tend to follow code sequentially, whereas experts shift across regions based on control flow and evolving hypotheses~\cite{icpc20-drives, peachock2017investigating, icpc15-relax}.

Because these measures can indicate different cognitive states rather than a single, directly observable construct, eye-tracking data is often interpreted together with interviews, surveys, or task performance measures~\cite{peachock2017investigating}.
Importantly, prior work also shows that physiological differences do not necessarily imply performance differences.
\citet{icer23-styling} found that background styling in code editors altered novices’ gaze behavior without significantly changing code comprehension performance.
Similarly, other work reported eye-movement differences for program comprehension without a significant difference in response time~\cite{peachock2017investigating} or accuracy~\cite{sbes20-atoms}.
However, prior eye-tracking studies have focused on attention patterns on statically presented code \cite{icpc12-women, icpc20-drives}. 
Much less is known about how programmers engage with structural representations such as ASTs \cite{tsem24-tale}, or how their attention changes when code is revealed dynamically. 
To address this gap, we used a controlled experimental setup that isolates rendering as the primary variable, allowing us to examine how different presentation strategies shape visual attention independent of higher-level task strategies or performance differences.

\subsection{AI-Resilient Interfaces}
AI-generated outputs are often difficult to verify because they can be plausible even when wrong, leading users to accept them without careful verification~\cite{25-hallucination, kalai2025languagemodelshallucinate}.
This challenge is exacerbated by interface designs that encourage users to skim rather than critically assess model responses.
In programming contexts, the problem is more acute because generated code is information-dense, compresses substantial logic into compact expressions, and can easily overwhelm users’ attention and working memory~\cite{icse24-ai-programming, uist24-steer}.
For example, prior work shows that learners may skim AI-generated responses and develop only an illusion of understanding~\cite{iui25-exploring}.
Rapid suggestions and long responses can intensify this pressure, further reducing users’ focus, leaving little time for critical assessment or internalization of code logic~\cite{uist24-steer, icer24-gap, chi25-tradeoffs}.

To address these challenges, prior work frames \say{AI-resilient} interfaces as designs that move users from post hoc error recovery toward earlier detection and assessment of AI outputs~\cite{chi24-ai-resilient}.
Related work similarly emphasizes proactive interface strategies that guide attention and support deeper processing, helping users resist plausible but incorrect responses~\cite{iui20-proxy}. 
Examples include saliency-based presentations such as GP-TSM, highlighting potentially important content for inspection~\cite{chi24-ai-resilient}, and adding intentional delays to encourage reflection rather than passive acceptance~\cite{chi24-ai-delays, su-etal-2023-reviewriter, cscw21-cognitive}.
Recent work explores cognitive-load-aware streaming, dynamically modulating the pace of LLM output based on inferred reading difficulty to better match user processing~\cite{uist25-streaming}.
In programming, one promising direction is to decompose code into semantic segments and deliver them incrementally~\cite{chi25-tradeoffs}.
Collectively, prior work highlights the importance of guiding users’ attention and supporting reflective processing in AI interactions. 
However, how generated content is revealed, particularly in dynamic settings, remains underexplored as a design dimension for shaping user attention and understanding.
\section{Structured Rendering}
\label{sec:design}
We define structured rendering as a presentation technique that progressively reveals code at the granularity of syntactic nodes in the program’s abstract syntax tree.

\subsection{Design Rationale}
Research in cognitive psychology demonstrates that experts chunk information hierarchically, processing high-level patterns before low-level details~\cite{miller1956magical}. 
However, the current model-centric code presentations misalign this cognitive preference by forcing linear, character-by-character streaming that obscures structural relationships.
We hypothesize that revealing code through its \ac{AST} hierarchy better aligns with programmers' mental models. 
Unlike left-to-right revelation, our approach prioritizes components by their logical dependencies, supporting the natural flow of program comprehension from control structure to implementation details.
By presenting syntactic skeletons first, the structural ``bones'' of constructs like loops and conditionals, we provide cognitive anchors that facilitate pattern recognition~\cite{sweller1988cognitive}. 
This structured approach ensures that programmers have the ability to continuously relate each piece of information, whether as a whole or in fragments, into a larger context~\cite{Degeng_2024}. 

\subsection{Illustrative Example}
To illustrate, consider the Java \texttt{for} loop in the \texttt{main} function (Figure~\ref{fig:render}, left), which iterates through integers and prints primes.
A typical linear reveal would obscure the loop's structure by presenting tokens in a flat sequence. 
Our approach (Figure~\ref{fig:render}, right, step 1–17), however, begins by emitting the syntactic skeleton \texttt{for ()} so the control boundary is visible before any details. 
It then logically unfolds the header components: the initialization \texttt{int num = 2}, the condition \texttt{num <= limit}, and the update \texttt{num++}. 
After the header is established, the renderer produces the body skeleton \texttt{\{\}} (Figure~\ref{fig:render}, right, step 14) and subsequently renders the body’s contents, which in this example is a nested \texttt{if} statement.
This order ensures that each new detail is presented within an already-established syntactic context, directly supporting program comprehension.

\subsection{Formal Algorithm}
Formally, the algorithm takes three inputs (Algorithm \ref{alg:structured-rendering}): the program’s AST; an \textit{OrderRule} mapping each node type to an ordered list of component selectors (e.g., for a \texttt{ForStatement}, the order might be init, then condition, then update, then body); and a \textit{MakeSkeleton} function that, given a node and its ordered components, produces a compact placeholder view (for example, \texttt{for (...) \{...\} }for a \texttt{ForStatement}). The output is an ordered sequence of rendering steps, \textit{R}, which, when played back, yields a structured, readable reconstruction of the code.

Operationally, the traversal is depth-first. 
When \texttt{VISIT} encounters a token node (identifiers, operators, literals, and keywords treated as atomic leaves), it immediately appends the token’s text to \textit{R}. 
When it encounters a non-token node with a defined rule, it first calls \texttt{MakeSkeleton} to append the node’s skeleton to \textit{R}, thereby revealing the construct’s shape and boundaries. It then recursively visits components in the order specified by \textit{OrderRule} (defined in Table \ref{tab:order-rules}). 
For nodes without an explicit rule, \texttt{VISIT} descends through all children in their natural order. In all cases, the invariant is \say{skeleton first, components later} giving readers an early anchor for the control flow or expression context they are about to examine.

\begin{algorithm}[t]
\caption{Structured Rendering}
\label{alg:structured-rendering}
\DontPrintSemicolon
\KwIn{
  AST: Abstract Syntax Tree of the program;\\
  OrderRule: mapping from NodeType $\rightarrow$ ordered list of component selectors;\\
  MakeSkeleton: function that returns a placeholder view of a node (e.g., \texttt{for(...) \{...\}})
}
\KwOut{$R$: ordered list of rendering steps}
\SetKwFunction{Render}{STRUCTURED\_RENDER}
\SetKwFunction{Visit}{VISIT}
\SetKwFunction{IsToken}{IS\_TOKEN}
\SetKwFunction{TypeOf}{TYPE}
\SetKwFunction{Get}{GET}
\SetKwFunction{Children}{CHILDREN}

\Render{AST, OrderRule}:
\Begin{
  $R \gets [\,]$\;
  \Visit{AST.root}\;
  \Return{$R$}\;
}

\BlankLine
\Visit{node}:
\Begin{
  \If{\IsToken{node}}{
    append $node$ to $R$\; \tcp*{atomic token: identifier/operator/literal/keyword}
    \Return\;
  }
  $t \gets \TypeOf{node}$\;
  \If{$t \in$ \textit{dom}(\textit{OrderRule})}{
    $order \gets OrderRule[t]$\;
    $skel \gets$ \textit{MakeSkeleton}($node$, $order$)\;
    append $skel$ to $R$ \tcp*{skeleton-first reveal}
    \ForEach{$selector \in order$}{
      $c \gets$ \Get{$node$, $selector$}\;
      \If{$c \neq \varnothing$}{
        \Visit{$c$}\;
      }
    }
  }
  \Else{
    \ForEach{$child \in \Children{node}$}{
      \Visit{$child$}\;
    }
  }
}
\end{algorithm}

\begin{table}[h]
\centering
\caption{Example rendering rules for common \ac{AST} node types.}
\begin{tabular}{p{2.8cm} p{5cm}}
\toprule
\textbf{Node Type} & \textbf{Rendering Order (component selectors)} \\ \midrule
\texttt{ForStatement} & Skeleton $\rightarrow$ Initialization $\rightarrow$ Condition $\rightarrow$ Update $\rightarrow$ Body \\
\texttt{IfStatement} & Skeleton $\rightarrow$ Condition $\rightarrow$ ThenBlock $\rightarrow$ ElseBlock \\
\texttt{FunctionDeclaration} & Skeleton $\rightarrow$ ReturnType $\rightarrow$ Name $\rightarrow$ Parameters $\rightarrow$ Body \\
\texttt{VariableDeclaration} & Type $\rightarrow$ Identifier $\rightarrow$ AssignmentOperator $\rightarrow$ InitialValue \\
\texttt{ClassDeclaration} & Skeleton $\rightarrow$ Name $\rightarrow$ Superclass $\rightarrow$ Interfaces $\rightarrow$ Body \\
\texttt{MethodCall} & Callee $\rightarrow$ Arguments \\
\bottomrule
\end{tabular}
\label{tab:order-rules}
\end{table}

\section{Experimental Design}
\label{sec:exp}
We conducted a controlled laboratory study to examine how different rendering techniques affect programmers' visual attention and cognitive experience during code reading.

\subsection{Tasks and Stimuli}
We selected Java for its well-defined program structure, which supports \ac{AST}-based rendering, and widespread use in eye-tracking studies, enabling methodological comparability \cite{journal12-expertise, icpc12-women, icpc15-relax}.
Following prior work, we used three programs of comparable difficulty: $c_1$ \texttt{PrimeNumbers} (computing primes from $1$–$100$) \cite{icpc12-women}, $c_2$ \texttt{Factorial} (computing the factorial of an input) \cite{etra14-assessing}, and $c_3$ \texttt{Vehicle} (defining a \texttt{Vehicle} class with an \texttt{accelerate} method) \cite{icpc15-relax}. 
Consistent with eye-tracking practice, each program fit on a single screen to minimize navigation effects and ensure gaze reflected visual attention rather than interface interaction~\cite{icpc12-women, peachock2017investigating}.

To isolate rendering effect, each program was presented in one of the three styles: $r_1$ Static (full code visible at once), $r_2$ Character-based (incremental reveal with a typewriter-like effect), and $r_3$ Structured (our technique; Sec.~\ref{sec:design}).
A balanced Latin square ensured each participant saw each rendering exactly once, with program–rendering pairings and order counterbalanced across the sample.
To reduce expectation-driven reading strategies, we withheld program topics~\cite{an2013schema}, encouraging uniform initial comprehension based on the code itself. 
All participants were instructed to read with the goal of answering subsequent retention and comprehension questions.

\paragraph{Rendering Speed}
For dynamic conditions, we calibrated rendering speed to balance naturalistic AI output with human readability.
While GPT-5 models generate about 99.2 tokens per second~\footnote{https://artificialanalysis.ai/models/gpt-5-2}, typical silent reading speed of English text ranges from 175 to 300 words per minute (wpm)~\cite{19reading-rate}, corresponding to 200 to 340 ms per token.
We therefore applied this delay to enhance readability, which was validated in two pilot studies with participants possessing basic Java knowledge confirming that this speed allowed them to follow the code comfortably, neither too fast nor too slow.
Consequently, all three code snippets completed dynamic rendering within 30 seconds, followed by a four-second pause to avoid an abrupt ending and prepare participants for completion.

\paragraph{Apparatus}
The eye tracker we used was the SR Research EyeLink Portable Duo at 1000 Hz, positioned below a 24-inch monitor ($54.2 \times 30.5\,\mathrm{cm}$) with a resolution of $1920 \times 1080$ pixels. 
Participants were seated 60 cm from the monitor and 45 cm from the eye tracker, with head movements stabilized using a chin and forehead rest. 
In head-stabilized mode, the EyeLink has an average accuracy down to 0.15° and noise as low as 0.01° at 1000 Hz under optimal conditions\footnote{https://www.sr-research.com/eyelink-duo-technical-specifications/}.

\subsection{Participants}
We recruited 53 participants through campus posters and convenience sampling.
Eligibility required normal or corrected-to-normal vision and basic Java experience.
Given previous inconsistent definitions of programming expertise~\cite{sonnentag2006expertise}, we combined self-reports with follow-up interviews to classify participants.
\textit{Beginners} had experience limited to undergraduate Java coursework ($N=29$; 11 female, 18 male), while \textit{Intermediates} had additional development experience ($N=24$; 7 female, 17 male).
Most reported 3--5 years of programming experience, with some intermediates reporting 6 or more years.
The mean age was 23 years (SD=3.1).
All participants gave informed consent and received \$30 compensation for a one-hour session.

\subsection{Procedure}
This IRB-approved study used a head-stabilized eye-tracker in a laboratory setting.
Upon arrival, participants were briefed on the procedure and informed that the study examined how programmers comprehend AI-generated code, without disclosure of the rendering styles.
Each session began with a warm-up task, followed by three trials using different code snippets ($c_1$, $c_2$, $c_3$) each rendered in distinct styles ($r_1$, $r_2$, $r_3$). 
The trial order was counterbalanced using a Latin square design. 
Each trial began with eye-tracker calibration (limited to three attempts) followed by 34 seconds code viewing on a full-screen, non-interactive mode (no keyboard or mouse).
If they finished early, they notified the experimenter to stop tracking.
Immediately afterward, participants completed two quizzes:
(1) a retention quiz without code access, assessing high-level logic and syntax recall,
(2) a comprehension quiz with code access, adapted from \citet{icpc15-relax}, focusing on semantic understanding.
Each quiz included three multiple-choice questions, three true/false questions, and one fill-in-the-blank item.
Sessions concluded with a semi-structured interview regarding AI experience and rendering perceptions.
Data collected included raw gaze recordings, quiz scores, and interview responses.
\section{\sys{} Dataset and Analysis}
\begin{figure*}[!h]
    \centering
    \includegraphics[width=1.01\linewidth]{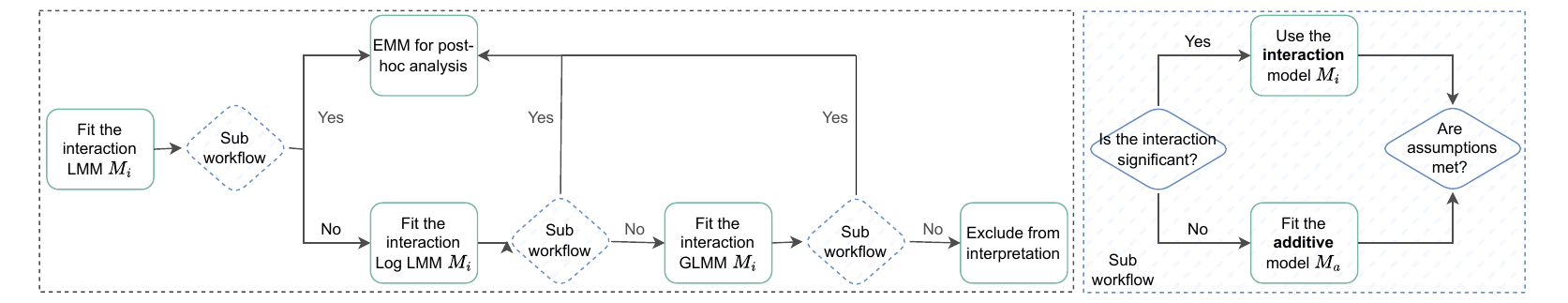}{
        \caption{This flowchart outlines a decision workflow for fitting and evaluating the mixed-effect models. The process begins by fitting a LMM that includes an interaction term $M_i$. The sub-workflow, shown in the green dashed box on the right, assesses each model for both the significance of the interaction and whether it meets at least four of five key assumptions. If the interaction is significant and assumptions are met, the interaction model is used; otherwise, an additive model $M_a$ is fitted. If this model also fails the assumption check, the process branches out to explore alternative model types, such as the Log LMM or GLMM. If no model can satisfy the assumption criteria, the variable is ultimately excluded from interpretation.}
        \Description{Flowchart showing decision tree for statistical model selection. Diamond-shaped decision nodes test interaction significance and assumption validation, with rectangular process boxes for fitting Linear Mixed Models (LMM), Log LMM, and Generalized Linear Mixed Models (GLMM). Green dashed box highlights assumption checking workflow requiring four of five criteria.}
        \label{fig:models}
    }
\end{figure*}
To support reproducibility, we release the study dataset (\sys{}) along with the analysis pipeline as open-source resources.
We designed the analysis pipeline to ensure reliable eye-tracking measures and robust inference across repeated-measures conditions. 
The process consisted of four steps: (1) verifying data quality, (2) preprocessing and mapping fixations to \acp{AoI}, (3) grouping dependent variables in two metric classes, and (4) modeling outcomes using linear and generalized linear mixed-effects models.

\subsection{Data Quality}
We first assessed calibration quality using average and maximum validation errors (Table~\ref{tab:calibration}). 
To further verify reliability, we examined the \emph{main sequence}---the relationship between saccade amplitude and peak velocity~\cite{DISTASI2011807,bahill1975main}. 
The observed trend followed the expected approximately linear pattern~\cite{bahill1975main}, confirming the integrity of the eye-tracking signal (Appendix~\ref{appendix:main-seq}). 
We excluded eight participants due to incomplete gaze data caused by early apparatus adjustments, and one additional participant for yielding fewer than 60 valid fixations per session (vs.\ $>$100 for others) after fixation mapping.
This resulted in a final analytic sample of $N=44$ for gaze analysis. As all participants completed the same procedure, the full sample ($N=53$) was retained for qualitative analysis.
Finally, we assessed video frame dropouts under dynamic rendering conditions.
Dropouts were rare, occurred primarily at trial onset, and resulted only in the first 2--3 words appearing simultaneously; playback thereafter was smooth, with no duplicated or missing content.
\begin{table}[h!]
\centering
\caption{Validation-error statistics (mean $\pm$ standard deviation) for the left and right eye computed across all $N=53$ participants prior to exclusion. Errors are reported in degrees ($^\circ$).}
\begin{tabular}{ccc}
\toprule
\textbf{Metric} & \textbf{Left Eye} & \textbf{Right Eye} \\
\midrule
Average Validation Error ($^\circ$) & $0.42 \pm 0.13$ & $0.44 \pm 0.15$ \\
Maximum Validation Error ($^\circ$) & $1.15 \pm 0.69$ & $1.10 \pm 0.65$ \\
\bottomrule
\end{tabular}
\label{tab:calibration}
\end{table}

\subsection{Event Detection and AoI Mapping}
Raw gaze data was exported from the EyeLink system and processed using \texttt{pymovements}~\cite{ETRA23-pymovements}.
We applied the Velocity-Threshold Identification (I-VT) algorithm~\cite{salvucci2000identifying}, classifying samples with velocities exceeding $20^\circ/\mathrm{s}$~\cite{sen1984effects}, and grouping the remaining samples into fixations with a minimum duration of $100$~ms.
These thresholds align with established definitions of fixations as spatially stable periods (typically $100$--$300$~ms~\cite{sharafi2020practical}), and saccades as rapid transitions between them~\cite{ist15-systematic,rayner1998eye}.

Fixations were mapped to \acp{AoI} corresponding to regions of interest in the code display.
For dynamic conditions, this required both spatial and temporal annotation: bounding boxes were marked in DataViewer\footnote{\url{https://www.sr-research.com/data-viewer/}}, and timestamps were identified using a custom annotation tool that aligned video playback with the gaze and fixation data at a millisecond resolution.
\acp{AoI} were defined at the point when a full word became visible, ensuring consistency across rendering conditions.
Each fixation was assigned to the nearest visible AoI containing the fixation point.
If fewer than 60\% of a participant's fixations were mapped, bounding boxes were expanded by 10 pixels in all directions. 
Across the final sample ($N=44$), the mean mapping success rate was 86.7\%.

\subsection{Dependent Variables}
We categorized dependent variables in two metric classes to examine developers' attention allocation and code inspection strategies. 

\subsubsection{Attention Metrics}
These metrics capture the intensity of visual attention during code comprehension. We measured \textbf{Fixation Rate}, defined as the number of fixations divided by the total view time \cite{andrzejewska2016comparing} and \textbf{Fixation Mean Duration}, which reflects the amount of time spent in a location; longer fixations indicate that participants spent more time analyzing and interpreting the content of the \acp{AoI}. 
The \textbf{Fixation Duration Standard Deviation} is stable within individuals over time and thus reflects reliable underlying cognitive processes \cite{henderson2014stable, staub2013individual}.
Finally, \textbf{Saccade Length} is the average Euclidean distance between consecutive fixations within a trial; prior work often finds that experts make longer saccades than novices \cite{icpc15-relax}.

\subsubsection{Linearity Metrics}
These metrics characterize the sequential structure of gaze during code reading. 
We classify saccades as forward (following a linear order) or backward (regressions). 
The \textbf{forward-saccade} metric is the percentage of fixations preceded by a forward move, comprising horizontal later (rightward within a line), vertical next (down to the next line), and vertical later (down two or more lines). 
The \textbf{backward-saccade} metric is the percentage of fixations preceded by a move against this order. 
Backward moves are further divided into global regressions (to an earlier line or block) and line regressions (leftward within the same line). 
Formal definitions and illustrations are in Appendix \ref{appendix:linearity}.

\subsection{Statistical Modeling}
\begin{table*}[h!]
  \centering
  \caption{The result for two metric classes: attention and linearity. The rendering condition was included using centered (sum-to-zero) Helmert contrasts: 
  (1) Statics vs Dynamic: A positive coefficient (Estimate) indicates the Static mean is higher than the average of Character and Structured. 
  (2) Structured vs Character: A positive coefficient indicates the Structured mean is higher than the Character mean. 
  Asterisks indicate significance (Sig.) (*$p<0.05$, **$p<0.01$, ***$p<0.001$). 
  Cohen's d represents the standardized effect size for each contrast. 
  All model estimates and effect sizes are reported with 95\% confidence intervals.}
  \begin{tabular}{
    l 
    S[table-format=+1.3] 
    S[table-format=1.3] 
    l                   
    l                   
    S[table-format=+1.3] 
    S[table-format=1.3] 
    l                   
    l                   
    l 
  }
  \toprule
   & \multicolumn{4}{c}{\textbf{Static vs Dynamic}} & \multicolumn{4}{c}{\textbf{Structured vs Character}} & \\
  \cmidrule(lr){2-5}\cmidrule(lr){6-9}
  \textbf{Measurement} & {Estimate} & {Cohen's $d$} & {$p$} & {Sig.} & {Estimate} & {Cohen's $d$} & {$p$} & {Sig.} & \textbf{Model} \\
  \midrule
    Fixation Rate              &  0.027341 & 4.770 & $< .001$ & ***  & -0.025776 & 2.595 & 0.009   & ** & Gamma GLMM \\
    Fixation Mean Duration     & -0.027872 & 4.952 & $< .001$ & ***  &  0.026878 & 2.757 & 0.006   & ** & Gamma GLMM \\
    Fixation Duration SD       & -0.05865  & 5.049 & $< .001$ & ***  & 0.08121   & 4.036 & $< .001$  & *** & Log-LMM \\
    Saccade Length             &  0.040506 & 5.068 & $< .001$ & ***  & 0.008062  & 0.583 & 0.561 & & Log-LMM \\
  \midrule
    Vertical Next              & -0.10777  & 5.406 & $< .001$ & ***  & -0.02386  & 0.658 & 0.511 & & Beta GLMM \\
    Vertical Later             & -0.014799 & 5.286 & $< .001$ & ***  & -0.004863 & 1.004 & 0.318 & & LMM \\
    Horizontal Later           & -0.028227 & 6.246 & $< .001$ & ***  & -0.004792 & 0.613 & 0.542 & & LMM \\
    Global Regression          &  0.015930 & 4.515 & $< .001$ & ***  & -0.000683 & 0.112 & 0.911 & & LMM \\
    Line Regression            &  0.01782  & 0.869 & 0.385 & & -0.05161 & 1.448 & 0.148 & & Beta GLMM \\
  \bottomrule
  \end{tabular}
  \label{tab:results}
\end{table*}

\acp{LMM} handle missing and unbalanced data more effectively, avoid strict sphericity assumptions by directly modeling covariance structures, and support random effects and continuous covariates. 
These properties make \acp{LMM} more flexible and robust than mixed ANOVAs, which require balanced designs and complete data. 
Therefore, we used \ac{LMM} and \ac{GLMM} to analyze the dependent variables.
We fitted mixed-effects models with fixed effects for rendering, expertise, their interaction, and random intercepts for participant and snippet to control baseline differences. 
The initial specification was therefore:
\begin{equation}
y \sim \text{render} \ast \text{expertise} + (1 \mid \text{participant}) + (1 \mid \text{code})  
\end{equation}

If the interaction was not statistically significant, we simplified the model by removing the interaction and retaining only the additive fixed effects.
We applied Helmert contrasts to the three-level rendering factor to test two hypotheses: (1) whether the Static condition differed from the average of the two Dynamic conditions (Structured and Character), and (2) whether Structured differed from Character-based rendering within the Dynamic conditions.
Full model specifications and validation are provided in Appendix \ref{appendix:stats}.
We first fitted the model, then we run post-hoc analysis and pairwise comparison.
Figure~\ref{fig:models} demonstrates the model fitting and selection process.

\subsection{Qualitative Analysis}
Additionally, we collected interview data to capture participants’ experiences of viewing code presented in three ways and to triangulate the eye-tracking results.
We conducted a thematic analysis following a consensus-coding protocol \cite{mcdonald2019reliability}. 
First, two researchers independently coded ten participant transcripts, extracting short phrases and key terms, using verbatim quotes as examples. 
We then merged our codes and discussed discrepancies in our choices and definitions. This process led to the creation of a preliminary codebook. 
\section{Findings}
\begin{figure*}
    \centering
    \includegraphics[width=1.05\linewidth]{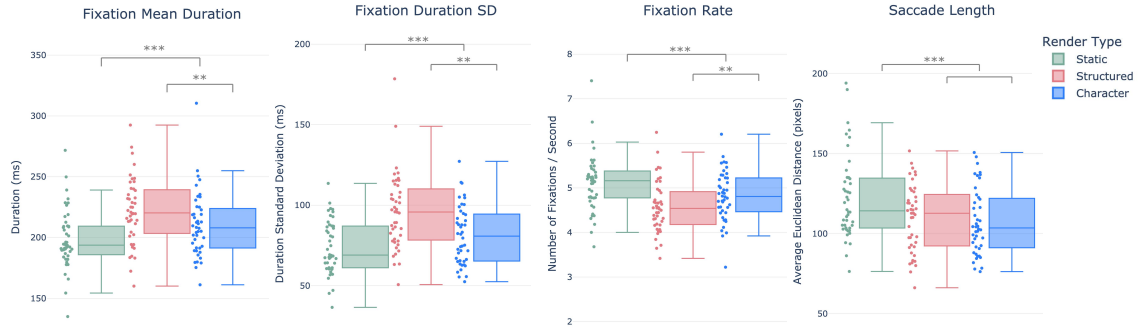}
    \caption{Boxplots showing eye-tracking metrics for the attention class across three rendering conditions (Static, Structured, and Character). Statistical significance is indicated by brackets above the plots (*$p<0.05$, **$p<0.01$, ***$p<0.001$).}
    \Description{Four boxplots comparing attention metrics between Static, Structured, and Character rendering conditions. Shows Fixation Frequency (fixations per second), Fixation Mean Duration (milliseconds), Fixation Duration Standard Deviation, and Saccade Length (pixels). Statistical significance brackets indicate Static condition has higher fixation frequency but shorter durations than dynamic conditions.}
    \label{fig:metrics}
\end{figure*}

Table~\ref{tab:results} summarizes the key statistical results for our main outcome metrics, highlighting significant effects across conditions. Notably, static and dynamic rendering conditions show distinct patterns in attention allocation (see Figure~\ref{fig:metrics}) and reading behavior, while structured and character-based rendering differ in their impact on the cognitive process of code comprehension and user experience.
We organize our findings along the two dimensions: (1) static versus dynamic rendering, and (2) structured versus character-based dynamic rendering. For each, we present quantitative results followed by qualitative insights to triangulate findings and provide a comprehensive understanding of participant behavior and perceptions.

\subsection{RQ1: Static vs. Dynamic}
We found that static and dynamic rendering evoked fundamentally different attention allocation patterns and reading strategies. In the static condition, participants exhibited a significantly higher fixation rate ($p < .001$), but each fixation was shorter on average ($p < .001$). Fixation durations also showed less variability ($p < .001$), and participants made longer saccades ($p < .001$) in the static condition. Moreover, static rendering encouraged less linear reading: participants were significantly less likely to progress vertically to the next line ($p < .001$) and more likely to make regressions, both globally ($p < .001$) and within lines.

\subsubsection{The Attention Patterns: Broad Scanning vs. Detailed Reading}
Our results show clear differences in how static and dynamic rendering shaped participants’ attention allocation and engagement with the code, even accounting for the fact that dynamic rendering necessarily imposes a more linear viewing order. 
The fixation and saccade metrics reveal how attention is distributed under each condition, offering insight into users’ broad scanning versus deeper, detail-oriented reading.
The static condition produced more frequent but shorter fixations, reflecting faster global scanning and more distributed but potentially superficial attention. 
This pattern was reinforced behaviorally: seven participants ended static sessions early, feeling ready to move on without using the full time.
In the dynamic condition, participants made fewer fixations overall but spent more time on each one, indicating sustained focus on individual code elements. 
Dynamic rendering also showed larger variance in fixation duration, suggesting that attention was allocated less uniformly ($p < .001$).
All participants continued until the end in the dynamic condition, indicating that they either remained engaged or felt compelled to follow the code as it unfolded. 

Qualitative reports echo these findings. 
In the static condition, participants favored having the complete code visible from the outset so that they are free to jump around and do not need to wait for the generation.
P50 wanted \psay{a big picture of the code,} so that they could \psay{read and think} in their usual way (P2, P34). 
Five participants (P11, P13, P23, P37, P50) explicitly described the static condition as offering greater flexibility in reading anything and control over their reading strategy because they could \psay{jump lines.} (P13) and \psay{skip and scan} (P37).
However, this sometimes reduced attention to detail; P26 reported that they \psay{jump around more and notice details less}.

In contrast, participants consistently reported that dynamic rendering promoted deliberate reading and attention to detail.
P8 noted that it \psay{forces you to read every part of the code}, a view echoed by six other participants (P11, P26, P34, P40, P51, P53). 
For example, P26 explained that dynamic rendering gave them \psay{less of an overlook}, while P51 said it \psay{kept me attentive}. 
P33 added that it \psay{helped with staying focused on one part}.
At the same time, this attentional guidance was not necessarily uniform. 
P17 noted that (in dynamic rendering) seeing the \psay{high level AST} made them \psay{a bit more selective}, suggesting that structured unfolding can guide readers' focus.
Some participants linked this attentional guidance to an active reconstruction process. 
P47 explained, \psay{my brain is reconstructing the code as it generates}, yielding \psay{more confident} understanding compared to static code. 
This suggests that observing structured generation can prompt parallel mental construction that strengthens comprehension.                        
The increased attention also supported recall and engagement.
P46 reported \psay{Dynamic rendering helped me remember more detail}, while P47 emphasized that it made them \psay{more engaged with the code and its purpose}. 
P20 similarly noted remembering the code better because they recalled \psay{how it was generated}.

\subsubsection{The Navigation Trade-off: Programmer Agency vs. Structured Guidance}
While static presentation affords freedom and control, dynamic rendering can support readers with unfolding information flow and sustaining focus.
Several reported feeling overwhelmed and uncertain about where to begin when viewing static code e.g., \textit{\say{For static code I look around and try to figure out where I should focus}} (P26, echoed by P20, P21, P23-P25). 
By contrast, many found the dynamic condition \psay{easier to follow} (P27, P36) with the unfolding animation \psay{naturally} guiding attention (P20, P25, P31) and reducing the effort of deciding where to look and in what order to read.
This guided focus also supported comprehension; as P26 explained, \textit{\say{In the dynamic case, it’s more step by step, and then it helps me to overlook everything again to understand how the concepts tie together.}}
Similarly, P31 remarked that static presentation \textit{\say{risks your eyes going somewhere else,}} whereas motion ``automatically'' draws focus.

\subsection{RQ2: Structured vs. Character}
Structured and character-based rendering also produced distinct visual attention patterns. 
In the structured condition, participants exhibited a lower fixation rate ($p < .01$), longer fixation durations ($p < .01$), and greater variability in fixation duration ($p < .001$,) compared to the character-based condition. 
Participants in this condition also tended to make longer saccades, although this effect was not significant. 
Linearity measures did not show significant differences between structured and character renderings.
Despite these attentional differences, performance outcomes were similar across the two conditions: retention scores averaged 4.36 in the structured condition and 4.43 in the character condition, while comprehension scores averaged 5.00 and 4.93, respectively (Appendix \ref{appendix:quiz}).

\subsubsection{From Distraction to Direction: How Meaningful Chunks Channel Focus}
Participants consistently found structured rendering more supportive for comprehension because it presented code in meaningful units rather than isolated fragments, while also foregrounding the overall structure of the code. 
Several noted that seeing complete syntactic elements, such as keywords, words, brackets, and function-level structure, helped them orient more quickly, sustain attention, and build an early sense of scope. 
P1 described viewing \psay{one syntactic unit at a time} as \psay{really helpful}, while P21 contrasted this with character-by-character rendering, which felt \psay{all over the place} and made it harder to stay focused. 

Beyond supporting local comprehension, participants also valued how structured rendering foregrounded scope before details. 
By presenting structural elements such as brackets, function definitions, and classes early, it provided a clearer sense of the code’s organization and high-level layout. 
Six participants (P16, P17, P18, P21, P27, and P51) highlighted that structured rendering made it easier to see the \say{scope} and overall \say{structure} of the code. 
P51 described the structured version as \psay{more holistic} and said it was \psay{nice to see the scope right away}. 
P18 described naturally searching for a closing bracket after encountering an opening one, while P21 said it was \textit{\say{easier to differentiate}} components such as functions, classes, and modifiers.
Similarly, P16 noted that having \psay{the whole structure first} made it easier to follow the function, while P21 said it was \psay{easier to differentiate} components such as functions, classes, and modifiers. Others likewise emphasized that structured rendering supported a stronger \psay{high-level overview} and a better sense of \psay{how the code is organized} (P17, P27, P35, P49).
Participants further linked these benefits to efficiency and syntax cues. 
Seeing complete words and meaningful structural units reduced guesswork and made the unfolding code easier to interpret.
As P43 explained, seeing \psay{the whole word} helped them \psay{focus and reduce the time to understand} because they no longer had to anticipate the next character.

In contrast, character-based rendering was perceived as fragmented, unnatural, and low in informational value. Participants described unfinished substrings and \psay{half finished words} as offering little meaning until completion, making it harder to retain the overall structure of the code (P43, P45). Some also found the fine-grained pacing frustrating, especially when it slowed down boilerplate content or presented code in ways that did not match natural programming practice (P1, P18, P19). 
These accounts suggest that granularity is not merely a presentation detail: rendering code in meaningful chunks better aligns with how programmers parse syntax, maintain attention, and build structural understanding.

\subsubsection{The Perceived Naturalness of Rendering Styles} 
Several participants associated structured rendering with natural coding practices.
P14 found it \textit{\say{comfortable because that’s like also how we understand code,}} while P47 described it as \textit{\say{logically more consistent with the way I will be doing it,}} and P19 said it \textit{\say{felt more like actually watching someone write code.}} 
However, some participants noted challenges due to unfamiliarity.
P15 noted \psay{I’m not used to it...it caught me off guard, but you could get used to it}, suggesting an adjustment period before its benefits become clear.
In contrast, some participants described character-by-character streaming as more natural, fluent, and intuitively aligned with how they read natural language, which made the unfolding code feel smooth and engaging (P14, P47, P51, P53). 
As P14 explained, it felt \psay{more natural and easier to read because it was more fluent}, while P47 called it \psay{more intuitive for how we use natural language...in a very linear order}.

\subsubsection{Beyond the Lab: How Programmers Experience Dynamic Rendering}
In our interview, we asked about participants' prior experience viewing AI-generated code that appears dynamically.

\paragraph{The Goldilocks Problem: When AI Generates Too Fast or Too Slow}
Participants described a persistent mismatch between AI rendering speed and their cognitive processing.
Inappropriate pacing disrupted concentration and created cognitive bottlenecks, leading 60\% of participants to wait until rendering finished before reading.
When too fast, the output outpaced reasoning and short-term memory, prompting attempts to \psay{keep up} that often sacrificed understanding, required restarting, or led participants to give up and wait (reported by ten participants). 
P13 noted having \psay{less time to understand the parts I cared about}, while P16 explained that when rendering is \psay{so quick ... I still need some time to process}. 
P40 added \psay{while I was still thinking, it had already written the rest}. 
As a result, some participants \psay{started again} (P19) or \psay{give up on following} (P45).

Participants also linked speed to memory and depth of processing.
Extremely rapid emission like ChatGPT was seen as unreadable in real time: \psay{it will spit out 50 lines of code in less than a second...better for it to fully render and then read it} (P52).  
At such speeds, participants felt they \psay{cannot remember much} (P46) and end up \psay{trying to keep up...not actually processing the code} (P48).
Conversely, when rendering was too slow, enforced waiting led to boredom, irritation, and disengagement, as noted by P13 and P45.
Even when watching the output, some did not actively process it, explaining that they \psay{let it finish} (P29). 
Others emphasized the need for control, preferring to \psay{read at my own speed} (P20) and \psay{ignore how fast it generates} (P21).

\paragraph{Task Switching and Distractions Under Delay.}
Faced with delays, participants often drifted into counterproductive strategies. 
They either waited passively and risked idleness or switched tasks and incurred context-switching costs.
As P10 noted, they were forced to \psay{wait through the boring parts}. 
Others described losing focus to distractions: \psay{nearly all the time I’m just switching to something else, another tab or looking at a paper} (P16), which they recognized as harmful: \psay{staying focused on one task is a lot better} (P16). 
Participants reported toggling between tabs (P36, P43), using the wait as a break (P35), or only multitasking when prompts were complex (P41). 
\section{Discussion}
\subsection{From Model-Centric to Human-Centric Rendering}
Our findings show that code presentation is not neutral: the way code is rendered shapes developers' visual attention, code navigation, and cognitive experience.
Our structured rendering approach offers a first step toward human-centered code presentation.
It organizes rendering around code structure, encouraging readers to engage with semantically meaningful units. 
In doing so, it begins to shift code presentation from a model-centric logic to a human-centric one.

\subsubsection{User-Controlled Structured Rendering: A VSCode Prototype}
To demonstrate how human-centered rendering could move beyond fixed presentation modes, we developed a VSCode prototype that gives users direct control over how generated code unfolds.
In addition to automatic rendering, the prototype supports two intervention mechanisms. 
First, users can click Stop to immediately reveal the full output when continued waiting no longer supports understanding. 
Second, users can click Step to clear the current block and enter a manual stepping mode, in which the reveal process is controlled through the keyboard: pressing the right or down arrow advances to the next token, while pressing the left or up arrow reverts by one token. 
This design treats rendering not as a predetermined behavior of the system, but as an interactive layer that users can adapt to their own pace, attention, and inspection strategy. 
In doing so, the prototype illustrates a broader shift from model-centric delivery toward user-controlled code interaction, where developers can move fluidly between passive viewing, immediate access, and fine-grained stepwise inspection.

\subsubsection{Rendering as Attentional Scaffolding in Dynamic Code Interaction}
Dynamic rendering appeared to encourage more effortful, detail-oriented inspection, creating conditions in which readers may engage more deeply with local code structure and potentially become more likely to notice defects. 
We interpret this not as evidence that one format improves performance, but as evidence that rendering can scaffold different modes of inspection, some of which may be better suited to tasks that require careful scrutiny.

This has immediate implications for both learning and professional practice.
In educational settings, structured rendering could offer instructors an alternative to purely character-by-character live coding by revealing code in meaningful structural units rather than requiring it to be typed in real time. This could free instructors to focus on explanation and conceptual guidance, rather than having pacing implicitly dictated by typing speed~\cite{chi25-ea-stress}. 
For students, this benefit would translate into a code presentation that is easier to follow, as it exposes semantically meaningful units instead of either fragmenting the program into a character-by-character stream or overwhelming them with a complete snippet.
Educational interfaces could further extend this idea with personalized learning paths where rendering speed, granularity, and structural emphasis adapt to individual learning styles and expertise levels.
In professional settings, the same principle could support focused code review by revealing logical structure incrementally and directing attention toward critical regions, such as likely bug locations, security-sensitive logic, or performance bottlenecks.
Future systems might integrate automated analysis with structured rendering to highlight importance and interdependencies, transforming the interface into a collaborative partner that helps reviewers allocate cognitive resources more effectively.

A natural next step is to examine these effects in more demanding scenarios, such as debugging, code review, or program modification, where differences in attentional strategy may more directly translate into task outcomes. 
This broader question becomes even more important in agentic programming environments, where generated code, natural-language explanations, reasoning traces, and editor context compete simultaneously for developers’ attention. 
From this perspective, our study provides a foundational step: before designing interfaces for these increasingly dynamic settings, we first need to understand how rendering itself organizes attention in code reading.

\subsection{Design Implications: Toward Cognitively-Aligned Human-AI Interaction}
Our work connects to a broader emerging view of generative interfaces.
Recent work on cognitive-load-aware streaming argues that LLM interfaces should adapt the pace of output to users’ reading demands rather than stream at rates determined only by computational throughput~\cite{uist25-streaming}.
Our findings suggest a complementary principle for code. 
In conversational text interfaces, the primary challenge may be temporal alignment: matching the speed of presentation to human reading and processing. 
In code reading, however, timing alone is not enough. 
Programmers do not simply consume output as it arrives; they seek signatures before details, infer hierarchy from syntax, skip predictable boilerplate, and organize information into meaningful chunks~\cite{icse14-automated, shneiderman1979syntactic, fse17-neural}. 
When interfaces do not account for these cognitive processes, more of the interpretive burden falls on the user.
Therefore, interfaces should not merely expose model output reflecting token generation. 
They should mediate output according to humans' cognitive demands.
\citet{uist25-streaming} advances this argument for pacing in text-based interaction; our work extends it to the structure and granularity of code presentation. 
These perspectives suggest that human-centered generation requires interfaces that adapt both the rhythm and the form of output to support understanding.

\subsubsection{Paradigm Shifts in Interface Design: From Static Presentation to Adaptive User-Controlled Interaction}
This broader view also points toward a new design agenda: adaptive user-controlled interaction for generative AI. 
Building on prior work showing that lengthy AI suggestions overwhelm users and interrupt their cognitive experience~\cite{chi23-grounded, chi25-tradeoffs, uist24-steer, uist25-streaming}, we found that overly fast rendering overwhelms processing and causes stress, while overly slow rendering disrupts flow and causes boredom.
Our structured rendering approach offers a first step to address this by synchronizing the reveal sequence with content structure, producing a more natural and cognitively compatible experience.
The next step is not simply to choose between static and dynamic presentation, or between character-based and structured rendering, but to design interfaces that can adapt pacing, granularity, and sequence to users, tasks, and content.
Readers differ in expertise, reading style, familiarity with the task, and tolerance for interruption.
Interfaces should therefore support interaction over rendering itself: pausing, stepping through meaningful structural units, expanding or collapsing details, and shifting between overview and detail as needed.
This vision of adaptive interactive rendering reframes generative presentation as a collaborative process between system and reader. 
Rather than asking users to continuously adapt to the output style of the model, future interfaces can help shape the output into forms that better fit human cognition and task demands.
More broadly, our findings contribute to ongoing HCI efforts to design AI interfaces that support critical engagement~\cite{chi24-ai-resilient} rather than passive acceptance as seen in \say{vibe coding}.

\subsubsection{Reduce Cognitive Costs of Generative Waiting}
Generative AI breaks established waiting paradigms. 
While the HCI community has agreed on a 10-second threshold for effective task focus \cite{93-usability, chi00-quality}, generative AI introduces unpredictable delays ranging from seconds to minutes \cite{chi25-ea-wait}, disrupting cognitive continuity in ways that traditional search does not.
During extended waits, users rely on limited working memory to sustain the continuity of complex mental models, increasing extraneous cognitive load: interruptions that demand effort without contributing to task progress~\cite{sweller1988cognitive}.
In programming, this leads to fading awareness of variable relationships and logic flow, hindering effective code integration and debugging.

Our study suggests that users respond to these delays in two costly ways: they either remain idle or switch to another task. 
Remaining idle wastes time, whereas task switching fragments attention and introduces resumption costs. 
Prior work shows that switching tasks leads to measurable lag in recovery~\cite{TRAFTON2003583}, and prolonged interruptions further increase recovery time and the risk of failing to resume the original task~\cite{chi17-resumption}. 
In programming, these disruptions can make it harder to reconnect generated output to the surrounding logic.

Structured rendering offers one way to reduce these costs by turning waiting into guided engagement.
Rather than forcing users to \psay{wait through the boring parts}, it uses generation time for scaffolding, which helps maintain cognitive continuity.
More broadly, this suggests that waiting in generative systems should be designed, not merely tolerated.
This challenge is especially important because generation is often slower and less predictable than retrieval, particularly for reasoning-heavy or multimodal tasks~\cite{chi25-ea-wait}. 
Beyond interface-level support, future systems could further reduce waiting costs through caching~\cite{uist25-streaming} or hybrid retrieval-generation strategies (e.g., Perplexity, Bing AI) to provide quick responses for familiar queries while reserving slower generation for complex tasks. 
Such strategies may help reduce both cognitive and computational costs~\cite{uist25-streaming} while preserving user focus during unavoidable delays.

\subsection{Limitations}
Our study encountered ceiling effects in traditional comprehension measures, with participants scoring near the maximum across conditions. 
Such quiz-based assessments emphasize outcomes over processes, missing key differences in cognitive effort revealed by our eye-tracking data. 
This highlights the need for process-sensitive measures that capture comprehension quality, cognitive fluency, and the sustainability of strategies across code complexities~\cite{wyrich202340}.

\subsubsection{Threats to validity}
\paragraph{Internal Validity.} 
Memory bias may affect internal validity because participants had to recall their experiences with different rendering styles. 
To mitigate this, we asked them to reflect immediately after completing each rendering quiz, grounding responses in a concrete experience. 
Misinterpretation of survey questions is another risk; we reduced this by piloting the survey twice and encouraging participants to ask clarifying questions.

\paragraph{External Validity.} 
Our controlled laboratory study, while methodologically rigorous, may not fully capture real-world programming practices. 
Because real-world reading goals vary (e.g., debugging, feature addition, refactoring), establishing this baseline under a standardized goal is a necessary first step; future work can examine rendering under specific real-world tasks.
\section{Conclusion}
Our eye-tracking study with 53 participants revealed that the rendering approach fundamentally shapes attention allocation and comprehension strategies. Static presentation enabled expert-like scanning with user control, while dynamic rendering provided attentional guidance that helped some participants manage cognitive load. Crucially, between two dynamic rendering techniques, structured rendering was favored over character-based approaches by presenting complete syntactic units that facilitated deeper processing and better structural awareness.
These findings suggest potential directions for AI-assisted programming tool design. Rather than mirroring the statistical generation process of language models, interfaces should align with human cognitive models of code comprehension. 
Structured rendering represents a promising step toward this alignment, offering a principled approach to progressive disclosure that leverages program structure to guide attention and enhance understanding. Future work could explore adaptive rendering systems that respond to individual expertise levels and task contexts, ultimately creating more effective human-AI collaboration across interactive systems.


\bibliographystyle{ACM-Reference-Format}
\bibliography{ref}
\newpage
\appendix

\section{Statistical Analysis Details}
\label{appendix:stats}
Participants $p$ are characterized by an expertise level $E_p$, which serves as a between-subjects factor. 
Each participant saw the three code stimuli, $c_1$, $c_2$, and $c_3$ (see Sec.~\ref{sec:exp}). 
Each snippet was rendered using one of three techniques — static ($r_1$), character-based ($r_2$), or structured ($r_3$). Let $R_{prc}\in\{r_1,r_2,r_3\}$ denote the rendering shown to participant $p$ for snippet $c$; rendering is treated as a within-subjects factor.
\subsection{Model Fitting Workflow}
Figure~\ref{fig:models} outlines a systematic approach to model selection and validation in mixed-effects modeling. 
We first fit the interaction model; if the interaction is not supported, we retain the additive model by setting $delta=0$. 
Next, we assess model adequacy using the unified five-assumption framework. If at least four assumptions hold ($|\mathcal{A}^*|\ge 4$), we proceed to post-hoc analysis. 
Otherwise, we refit the same specification with a logarithmic transformation \(\log(Y+c)\). 
If the transformed model still fails to meet adequacy criteria, we instead fit a \ac{GLMM} and reassess the five-assumption defined for GLMMs.

\subsubsection{Model Specification}
Formally, the full model combines fixed effects and random intercepts is defined as:
\begin{equation}
\begin{aligned}
\eta_{pr} &= \beta_0 + \beta_1 R_{prc} + \beta_2 E_p + \delta\,\beta_3 (R_r E_p) + u_{0p} + v_{0c},\\
\mu_{pr} &= g^{-1}(\eta_{pr}),\\
\tilde{Y}_{pr} \mid u_{0p}, v_{0c} &\sim \mathcal{F}(\mu_{pr}, \phi),
\end{aligned}
\end{equation}
The linear predictor $\eta_{pr}$ combines fixed effects and random intercepts. Here, $\beta_0$ is the intercept,$\beta_1$,$\beta_2$ capture the fixed effects of $R_r$ and $E_p$, while $\beta_3$ encodes their interaction. \(u_{0p}\) and \(v_{0c}\) are random intercepts for participant and code, respectively. The indicator \(\delta\in\{0,1\}\) toggles the interaction \((\delta=1:\mathcal{M}_i,\ \delta=0:\mathcal{M}_a)\); operationally, we begin with \(\delta=1\) and set \(\delta=0\) if the interaction is non-significant.

\subsubsection{Assumption-aware response definition.}
To promote robust inference, we adapt the modeled response according to a five-assumption audit. Let \(\mathcal{A}^*\subset\mathcal{A}\) denote the subset of assumptions that pass their diagnostic criteria. We use
\begin{equation}
\tilde{Y}_{pr} =
\begin{cases}
Y_{pr}, & \text{if } |\mathcal{A}^*| \ge 4,\\
\log\!\big(Y_{pr}+c\big), & \text{if } |\mathcal{A}^*| < 4,
\end{cases}
\qquad
c=\max\!\Bigl(0,\ \bigl|\min_{p,r}Y_{pr}\bigr|+1\Bigr),
\end{equation}
which ensures positivity and mitigates right skew and heteroscedasticity without altering the model structure.
If assumptions hold, post hoc analysis uses estimated marginal means (EMM) with Bonferroni-corrected pairwise tests; otherwise, the model is refit with a log-transformed outcome variable.

\subsubsection{Family–link selection and estimation.}
If log transformation is insufficient, we then fitted the GLMM with automated family selection based on data characteristics. The likelihood is chosen to match the support and dispersion of the data:
\begin{equation}
(\mathcal{F}, g, \phi) =
\begin{cases}
\mathcal{N}(\mu, \sigma^2), \text{ with } g=\text{identity},\ \phi\equiv\sigma^2 & \text{(Gaussian)}\\
\text{Beta}(\mu, \phi), \text{ with } g=\text{logit} & \text{(proportions)}\\
\text{Poisson}(\mu), \text{ with } g=\log,\ \phi\equiv 1 & \text{(counts)}\\
\text{Gamma}(\mu, \phi), \text{ with } g=\log & \text{(positive continuous)}
\end{cases}
\end{equation}
For count and binomial outcomes, overdispersion is evaluated using the Pearson \(\chi^2/\mathrm{df}\) ratio, with values in \([0.75,\,1.25]\) indicating adequate fit.
When \(\mathcal{F}=\mathcal{N}\) and \(g=\text{id}\), it is convenient to display
\begin{equation}
Y_{pr} = \eta_{pr} + \epsilon_{pr}, \qquad \epsilon_{pr}\sim \mathcal{N}(0,\sigma^2).
\end{equation}
making the residual variance explicit while leaving the fixed and random effects unchanged.

\subsubsection{Assumption Validation}
We implemented comprehensive assumption validations for both LMMs and GLMMs.
For LMMs, we assessed linearity, homoscedasticity, independence, and normality of residuals and random effects.
For GLMMs, we evaluated distributional and link function assumptions, independence, random effects normality, and potential overdispersion or zero-inflation.
Each category was examined using multiple diagnostics to ensure robust and consistent evaluation of model validity.

\subsubsection{Post-hoc Analysis}
For additive models, \acp{EMM} are computed independently for each main effect by averaging predictions across levels of other factors. 
For interaction models, they are estimated for all combinations of interacting factors, preserving joint effects.
Effect sizes are derived from fixed-effect contrasts as the absolute coefficient divided by its standard error, yielding standardized measures aligned with the tested contrasts.

\section{Quiz Results}
\label{appendix:quiz}
\begin{table}[h!]
\centering
\begin{tabular}{lcccc}
\toprule
\textbf{Score} & \textbf{Static} & \multicolumn{2}{c}{\textbf{Dynamic}} \\
\cmidrule(lr){3-4}
               &                 & Structured & Character \\
\midrule
Retention      & 4.659 & 4.364 & 4.432 \\
Comprehension  & 4.886 & 5.000 & 4.932 \\
\bottomrule
\end{tabular}
\caption{Mean retention and comprehension scores by render condition for $N=44$ participants.}
\label{tab:quiz}
\end{table}

\onecolumn
\section{Linearity Metrics}
\label{appendix:linearity}
\subsection{Formula Definitions}
\begin{table*}[htbp]
\centering
\renewcommand{\arraystretch}{2}
\begin{tabular}{lp{5.8cm}p{7.5cm}}
\toprule
\textbf{Metrics} & \textbf{Definition} & \textbf{Computation} \\
\midrule
Vertical Next & \% of forward saccades that either stay on the same line or move one line down. & $Y_{\text{VN}} = \frac{1}{N_{pr}-1} \sum_{i=1}^{N_{pr}-1} \mathbf{I}\!\big(L_{pr,i+1} - L_{pr,i} \in \{0, 1\}\big)$ \\
Vertical Later & \% of forward saccades that either stay on the same line or move down any number of lines. & $Y_{\text{VL}} = \frac{1}{N_{pr}-1} \sum_{i=1}^{N_{pr}-1} \mathbf{I}\!\big(L_{pr,i+1} \ge L_{pr,i}\big)$ \\
Horizontal Later & \% of forward saccades within a line. & $Y_{\text{HL}} = \frac{1}{N_{pr}-1} \sum_{i=1}^{N_{pr}-1} \mathbf{I}\!\big(L_{pr,i+1} = L_{pr,i} \land W_{pr,i+1} \ge W_{pr,i}\big)$ \\
Global Regression & \% of backward saccades of any length. & $Y_{\text{GR}} = \frac{1}{N_{pr}-1} \sum_{i=1}^{N_{pr}-1} \mathbf{I}\!\big((L_{pr,i+1} < L_{pr,i}) \lor (L_{pr,i+1} = L_{pr,i} \land W_{pr,i+1} < W_{pr,i})\big)$ \\
Line Regression & \% of backward saccades within a line. & $Y_{\text{LR}} = \frac{1}{N_{pr}-1} \sum_{i=1}^{N_{pr}-1} \mathbf{I}\!\big(L_{pr,i+1} = L_{pr,i} \land W_{pr,i+1} < W_{pr,i}\big)$ \\
\bottomrule
\end{tabular}
\caption{Linearity metrics in the $\mathcal{L}$-class for participant $p$ under render condition $r$. $N_{pr}$ denotes the total number of fixations, $L_{pr,i}$ represents the line number of fixation $i$, $W_{pr,i}$ is the within-line token index, $(x_{pr,i}, y_{pr,i})$ are screen coordinates, $\mathbf{I}(\cdot)$ is the indicator function, and $T_c$ represents the total number of tokens in code stimulus $c$. The operators $\land$ and $\lor$ denote logical AND and OR, respectively.}
\label{tab:linearity}
\end{table*}
\subsection{Visual Explanations}
\begin{figure}[h]
    \centering
    \includegraphics[width=\linewidth]{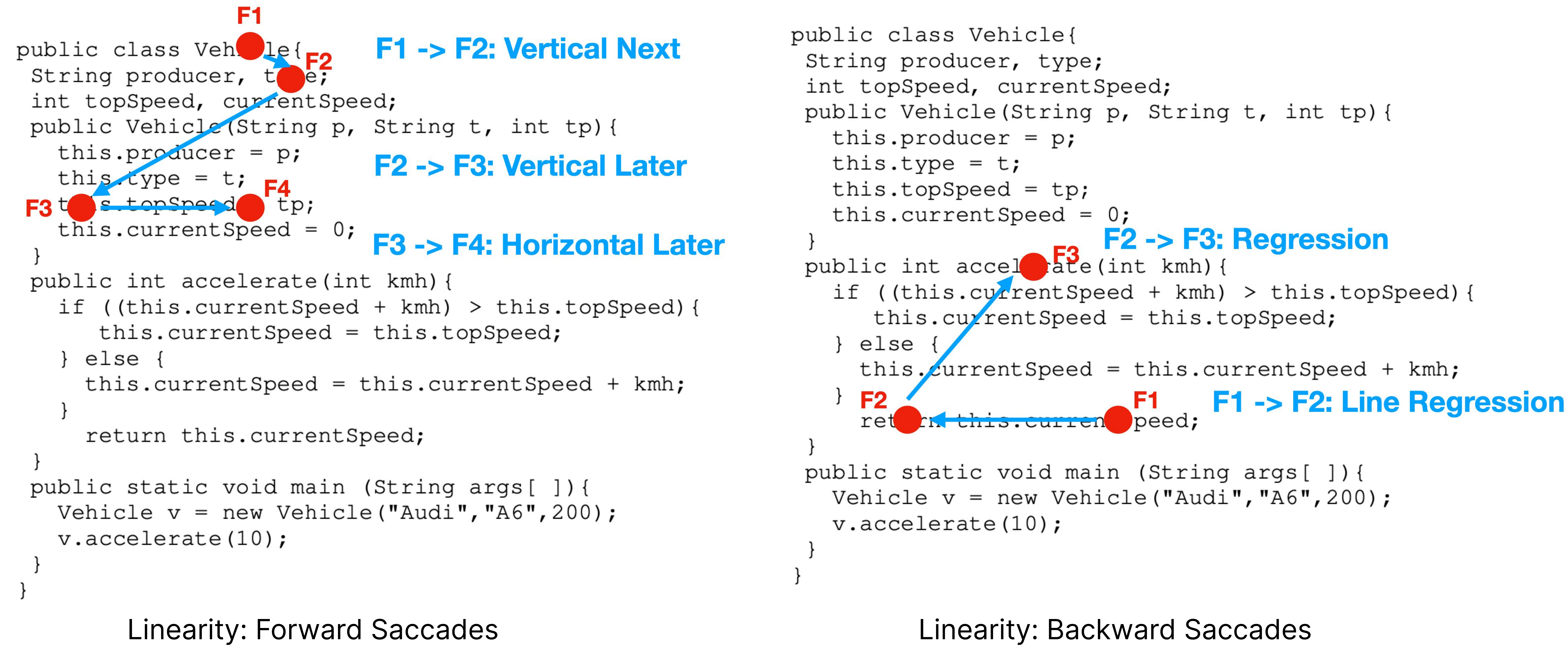}
    \caption{Linearity visual explanations}
    \Description{Diagram illustrating eye movement patterns during code reading. Shows arrows indicating forward saccades (horizontal later, vertical next, vertical later) and backward saccades (global regression, line regression) across lines of code text, with different arrow colors and styles representing different movement types.}
    \label{fig:linearity}
\end{figure}
\newpage
\section{Main Sequence Plot}
\label{appendix:main-seq} 
\begin{figure*}[htbp]
  \centering
  \begin{subfigure}[b]{0.45\textwidth}
    \centering
    \includegraphics[width=\textwidth]{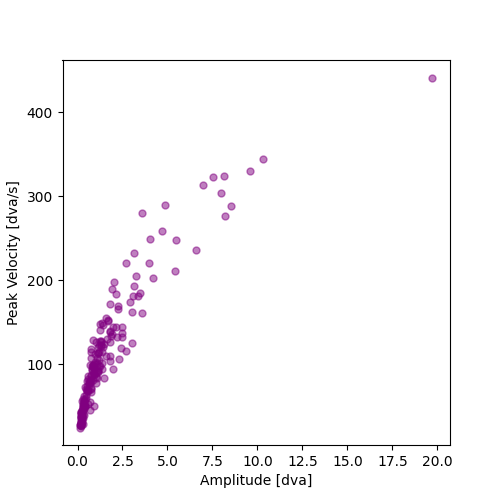}
    \caption{From P19}
  \end{subfigure}
  \hfill
  \begin{subfigure}[b]{0.45\textwidth}
    \centering
    \includegraphics[width=\textwidth]{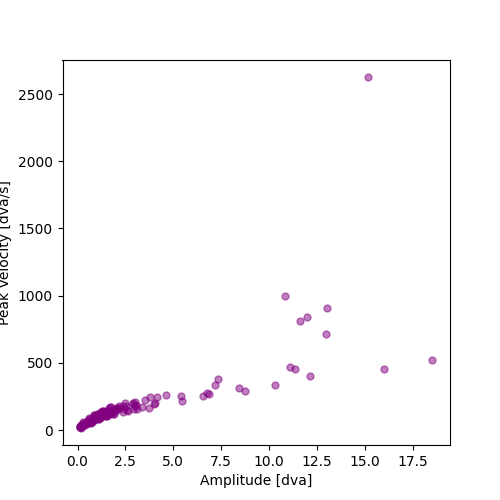}
    \caption{From P26}
  \end{subfigure}
  \caption{Main sequence plots sampled from two participant trials.}
  \Description{Two scatter plots showing saccade amplitude (degrees) on x-axis versus peak velocity (degrees per second) on y-axis for participants P19 and P26. Each plot displays positive linear relationship between saccade distance and velocity, with data points clustered along diagonal trend lines indicating normal eye-tracking calibration quality.}
  \label{fig:main-seq}
\end{figure*}


\end{document}